\documentclass{article}

\usepackage{arxiv}

\usepackage[utf8]{inputenc} 
\usepackage[T1]{fontenc}    
\usepackage{hyperref}       
\usepackage{url}            
\usepackage{booktabs}       
\usepackage{amsfonts}       
\usepackage{nicefrac}       
\usepackage{microtype}      
\usepackage{graphicx}
\usepackage{subfigure}
\title{The Risks of WebGL: Analysis, Evaluation and Detection}

\author{
  Alex Belkin\\
  Department of Electrical Engineering\\
  Technion University\\ 
  Haifa, Israel \\
  \texttt{belkinalex@gmail.com}
   \And
    Nethanel Gelernter \\
  Department of Computer Science\\
  College of Management Academic Studies\\
  Rishon LeZion, Israel \\
  \texttt{nethanel.gelernter@gmail.com } \\
     \And
  Israel Cidon\\
  Department of Electrical Engineering\\
  Technion University\\ 
  Haifa, Israel \\
  \texttt{cidon@ee.technion.ac.il }
}

\begin{document}
\maketitle

\begin{abstract}
WebGL is a browser feature that enables JavaScript-based control of the graphics processing unit (GPU) to render interactive 3D and 2D graphics, without the use of plug-ins. Exploiting WebGL for attacks will affect billions of users since browsers serve as the main interaction mechanism with the world wide web.
This paper explores the potential threats derived from the recent move by browsers from WebGL 1.0 to the more powerful WebGL 2.0.
We focus on two possible abuses of this feature: distributed password cracking and distributed cryptocurrency mining.
Our evaluation of the attacks also includes the practical aspects of successful attacks, such as stealthiness and user-experience. 
Considering the danger of WebGL abuse, as observed in the experiments, we designed and evaluated a proactive defense. We implemented a Chrome extension that proved itself effective in detecting and blocking WebGL.
We demonstrate in our experiments the major improvements of WebGL 2.0 over WebGL 1.0 both in performance and in convenience.
Furthermore, our results show that it is possible to use WebGL 2.0 in distributed attacks under real-world conditions. 
Although WebGL 2.0 shows similar hash rates as CPU-based techniques, WebGL 2.0 proved to be significantly harder to detect and has a lesser effect on user experience.

\end{abstract}

\section{Introduction}
\label{intro:sec}
The rapid evolution of web technologies has delivered new possibilities to billions of users, while at the same time exposing them to new security threats. Browsers are an excellent example of this phenomenon. Their new features are being abused by malicious hackers to create attack vectors and efficiently launch attacks not previously considered a risk.


Some recent examples are the Cache API~\cite{cacheapi} and the ServiceWorker~\cite{serviceWorker:w3c:2015} features, which allow the launch of sophisticated timing side-channel attacks \cite{van2015clock,advxssearch:bh16}. Another example is the Quota API, which can be used to extract the exact size of cross-site requests \cite{vanrequest}.

Browsers serve users as the main interaction mechanism with the world wide web. Therefore, security vulnerabilities or browser features that can be exploited for attacks will affect billions of users and close to two billion websites that are being accessed. Previous works by both Van Goethem, Gerlernter ~\cite{van2015clock,advxssearch:bh16,vanrequest}, and many others \cite{DBLP:conf/ndss/0001KK15,gelernter2015framing} offer examples of methods that exploit browser features to attack users. Other browser features, such as web-workers or web-sockets, have been used to effectively launch distributed denial-of-service attacks on websites \cite{erkkila2012websocket,pellegrino2015cashing,lam2006puppetnets}.

This paper explores the risks posed to web users by WebGL 2.0 \cite{webgl2:spec}.
Web Graphics Library (WebGL) is a JavaScript API that uses the graphics processing unit (GPU) to render interactive 3D and 2D graphics within any compatible web browser, without the use of plug-ins~\cite{webgl1}.
WebGL allows users to communicate directly with the graphics hardware. It comes with its own programming language called GLSL. GLSL allows anyone to control the computational power of the GPU, using it as they wish.

Initially, GPUs were designed to accelerate the creation of images intended for output to a display device. However, their highly parallel structure makes them more efficient than general-purpose CPUs for algorithms that process large blocks of data in parallel. This made GPUs rise to prominence in the fields of crypto-mining, deep-learning, and more. 

Controlling the GPU via the browser has introduced several new opportunities for attackers. For example, GPUs are used to efficiently break hashes~\cite{alcantara2009real} or for Bitcoin mining~\cite{taylor2013bitcoin}.
The ability to abuse the previous version of this API has already been examined by researchers. Fortunately, WebGL 1.0 is quite limited and does not support $32$-bit integers or bitwise operators ~\cite{webgl1:spec}. This makes it difficult to implement algorithms that efficiently calculate MD5 hashes, and much harder to implement more complex hashing algorithms such as SHA-2. Marc Blanchou’s presentation at Black Hat Europe ~\cite{GPU:blackhat13} showed that these limitations make the abuse of WebGL 1.0 ineffective, even compared to more naive implementations in JavaScript that use the CPU.

WebGL 2.0 was recently integrated into the popular browsers Google Chrome and Mozilla Firefox. WebGL’s new features (e.g., support for $32$-bit integers) establish the need to reevaluate the risks posed by this API. Due to the expected danger, it was also essential to develop countermeasures that can detect malicious WebGL 2.0 code and block it. This is a challenge that has not been studied before, even for WebGL 1.0.

Our initial research focused on distributed password cracking, where users' browsers are exploited to crack hashes while the user is surfing web pages controlled by the attacker.
In Section \ref{passwordcrack:evaluation}, we describe an experiment that enabled us to compare the effectiveness of password cracking using WebGL 1.0, WebGL 2.0, and CPU techniques. Our results show that although WebGL 1.0 demonstrates a slower hash rate than the CPU hasher, WebGL 2.0 shows significant improvement. Its hash rate is nearly two times faster than the CPU hasher when using best trade-off values as shown in Table 1.

A few months into the research, the crypto-currency market began its rapid growth, and our research on the abuse of browsers for hash cracking became a reality. 
Coinhive \cite{CoinHive} and other similar companies made it easy for every website owner to mine crypto-currency such as Monero \cite{Monero:Home}, on the browsers of people who visit their page. 
As a result, the browsers of hundreds of millions of web users were abused to mine crypto-currency. In most cases, it was done without the permission or the knowledge of the users \cite{CPUKiller:CheckPoint}.
This phenomenon encouraged researchers ~\cite{Cryptomining:BackShot,Cryptomining:Minesweeper} to perform an in-depth investigation of the landscape and impact of in-browser crypto-currency mining.

Both Hong et al. ~\cite{Cryptomining:BackShot} and   
Konoth et al. ~\cite{Cryptomining:Minesweeper} show how prevalent and potentially profitable crypto-currency mining can be for attackers. They explored the distribution of the infected websites containing mining code and demonstrate that no type of website is safe. Both works emphasize the inadequacy of current defense mechanisms, which are based on blacklists, and each suggested their own innovative countermeasure. Konoth et al. ~\cite{Cryptomining:Minesweeper} managed to identify as many as 20 different active crypto-currency mining campaigns in 0.18\% of Alex’s Top 1 Million websites.

Other researchers implemented a framework to allow persistent and stealthy bot operation through web browsers without the need to install any software on the client side called MarioNet~\cite{Cryptomining:Puppets}. The effectiveness of MarioNet is demonstrated by designing a large set of successful attack scenarios where the user’s system resources are abused to perform malicious actions including DDoS attacks to remote targets, cryptojacking, malicious/illegal data hosting, and darknet deployment.

At this point, we decided to add a new direction to our research and examine the consequences of WebGL 2.0 abuse for distributed crypto-mining. 

\subsection{Contributions}
The main contributions of this paper are the analysis and evaluation of the risks posed by WebGL 2.0 under real-world conditions, and a prevention method to WebGL 2.0 attacks. To the best of our knowledge this is the first work to analyze the risks associated with WebGL 2.0. We addressed both the computational and user experience aspects of such potential exploits. Specifically, our research studies the following questions:
\begin{enumerate}
\item Can WebGL 2.0 be used to launch practical attacks?
\item How effective is WebGL 2.0 for distributed password cracking and crypto-currency mining compared to WebGL 1.0 and CPU-based techniques?
\item What can be done to detect and block WebGL 2.0 attacks?
\end{enumerate}
Evaluating WebGL 2.0 only in terms of the theoretical attacker scenario benefit is not enough. It is also crucial to include different aspects that affect the effectiveness of distributed attacks under real-world conditions, such as user experience and stealthiness. Even if we manage to show a high performance attack under lab conditions, it isn't worth much to the attacker if the user can sense or even detect our attack. 
We implemented a distributed attack, which was used in several experiments on numerous users, to test all the necessary aspects of a distributed attack. 
Our results show that when it comes to performance, WebGL 2.0 and Coinhive show approximately the same hash rate as shown in Figure 2. However, for distributed attack aspects, our results in Section \ref{crypto_dist:sec} demonstrate that WebGL 2.0 is much harder to detect and has a lesser effect on user experience compared to CPU miner. Shown in Figure 3 and the results of Experiment 4, this proves that WebGL 2.0 miner is more suited to cryptocurrency mining distributed attacks than CPU miner.

We further implemented and tested a means to detect such an attack. In Section \ref {defenses:sec}, we implemented a Chrome extension to serve as a means for detecting and blocking the use of WebGL. We performed an experiment with numerous participants over an extended period of time to test the extension's efficiency and collect statistics about the use of WebGL in websites. Our results show that our extension is efficient in preventing WebGL abuse and that WebGL is relatively rare in websites.

Our findings are important and relevant to the Web community, mainly because the attacks studied in this paper have already been launched in different ways in the wild.

\subsection{Paper Organization}
Section \ref{preliminaries:sec} offers relevant background material about hash cracking, browser-based attacks, and the difference between GPU and CPU implementations.
Section \ref{related:sec} discusses related work. 
Sections \ref{passwordcracking:sec} and \ref{crypto:sec} analyze the abuse of WebGL 2.0 for password cracking and cryptocurrency mining, correspondingly. In both sections, the analysis is done from the perspective of a single victim and compares similar implementations using CPU and WebGL 1.0.
Section \ref{crypto_dist:sec} evaluates the abuse of WebGL 2.0 in a distributed attack under real-world conditions. 
Section \ref{defenses:sec} suggests and evaluates a means of defense, and Section \ref{conclusions:sec} concludes.

\section{Background and Motivation}
\label{preliminaries:sec}
The following section briefly explains concepts that will help the reader acquire a deeper understanding of the issues addressed in this paper. The section reviews hash cracking, browser-based distributed attacks, and the key differences between hash cracking implementation in WebGL and CPU.

\subsection{Hash Cracking} \label{PasswordCrack:sec}
This work addresses the challenge of hash cracking in two different scopes: password cracking and cryptocurrency mining.

Password cracking is the process of recovering passwords from the exposed output of a one-way cryptographic hash function, performed on the password. 
Password cracking can be done for several reasons, but the usual malicious reason is to gain unauthorized access without owner authorization or awareness.

There are dozens of password cracking programs on the market, each with its own special procedure\cite{Keeper:Security}. However, all usually do one or a combination of the following password searches: 
\begin{enumerate}
\item Create variations from a dictionary of known common passwords
\item Using brute force attack by trying all possible strings
\end{enumerate}

In cryptocurrency mining, hash cracking is needed to ensure the authenticity of the information and to update the blockchain with the transaction \cite{Blockgeeks:Hashing}.
Each time a cryptocurrency transaction is made, the cryptocurrency miners must solve complicated mathematical problems using cryptographic hash functions; these functions are associated with a block containing the transaction data. The mining process validates the calculated hashes on incremental values, called nonce, which are added to the given block data.  The hash output must match a certain criterion: it needs to be less than the cryptocurrency’s current target value. The current target value is also represented by the cryptocurrency’s difficulty. Cryptocurrency difficulty is a measure of how long would it take at a given hash rate to find a block that matches the current target. Higher difficulty means a lower target value. The difficulty increases over time and varies between cryptocurrencies.
Mining also involves competing with other cryptocurrency miners. Only the first one to crack the hash is rewarded with small amounts of cryptocurrency.

Advances in hardware and dedicated hash-cracking software have made hash cracking more practical and accessible than it used to be. For instance, a password's hash that would take over three years to crack in the year 2000 took just over two months to crack by the year 2016 \cite{BetterBuys:Password}.

One example of these hash cracking tools is an advanced password recovery tool created by Team Hashcat. Called Hashcat \cite{Hashcat}, this technology has won a succession of recent “Crack Me If You Can” contests \cite{CrackMe} and is described \cite{Hashcat,FossBytes} as the world's fastest password cracker. Hashcat enables any user to crack a significant number of different kinds of hash algorithms with ease and speed on multiple platforms, and introduces a variety of advanced features.
This led to designated hash cracking rigs that managed to break 300 GH/s on MD5 algorithm, 3493 KH/s on Scrypt algorithm, and more \cite{Github:Hashcat}.
Furthermore, hash cracking has been tested on cloud GPU systems (e.g., Amazon EC2), which are more accessible to users than dedicated rigs. These tests managed to reach a rate of 2494 MH/s on MD5 hashing algorithm on a single GPU \cite{Amazon:Hashcat}.

Browser-based CPU hash crackers written in JavaScript are generally slower than native CPU hash crackers due to their implementation and the browser's overhead. However, in recent years, a new browser-based CPU hash cracker was introduced: WebAssembly hash cracker.
WebAssembly (Wasm) is a binary instruction format for a stack-based virtual machine. Wasm is designed as a portable target for the compilation of high-level languages like C/C++/Rust, enabling deployment on the web for client and server applications \cite{WebAssembly}. Wasm is designed to be faster to parse than JavaScript, as well as faster to execute, enabling very compact code representation \cite{WebAssembly:Goals}.
This led to the introduction of Coinhive \cite{CoinHive}, the first browser-based CPU cryptocurrency miner.
Coinhive uses Wasm to increase the hash rate and reduce JavaScript overhead. 
As a result, the Coinhive Wasm miner outperforms JavaScript miners. Coinhive \cite{CoinHive} even states that they are able to reach about 65\% of the performance of a native miner.\newline
At the time of writing, we found no efficient browser-based GPU hash crackers.

\subsection{Browser-based Distributed Attacks}
Distributed computing takes complex computing tasks, such as breaking cryptographic hashes, and splits them up into smaller parts. It then sends them out to many different personal computers or servers to be processed in parallel and return with results. This parallel processing of many smaller parts serves to significantly reduce the time needed to compute each given task. In general, distributed computing uses abundant compute resources, including many CPUs, high network bandwidth, and a diverse set of IP addresses.

A browser-based distributed attack allows the attacker to exploit web users to perform distributed tasks at will. The attack starts when a victim enters a web page that is controlled by the attacker.
Opening a web page causes the web browser to initiate a series of background communication messages to fetch and display the requested page. This requires the client's web browser to download and run code that is served from the website on the client's device. The code that gets executed in the client's browser is assumed to be related to the functionality of the site being browsed. Technically, however, there is nothing stopping a website from serving arbitrary code that is not related to the browsing experience. With the absence of any protection, the client's web browser will blindly execute whatever code it downloads from the website. 

At this point, a malicious code can run and gain full access to the web browser's API, which presents an increasingly powerful set of web technologies. The code is transient and difficult to detect once the user has navigated away from the website. This gives the attacker access to the compute resources of all concurrent website visitors at any given time. This amount of compute power is especially significant on high-traffic websites. An attacker can take advantage of these opportunities to execute large-scale browser-based distributed attacks. For example, these attacks may exploit the victims' compute resources to perform in-browser distributed hash cracking.

In \cite{Dorsey:Medium}, Dorsey explains how easy it is to execute distributed attacks on the browsers of unsuspecting users. With small effort and funds (spending less than 100\$), he managed to reach thousands of victims, using paid advertisements as a means of distribution. This allowed him to freely run the code of his choosing in the clients' browsers. He then demonstrated the feasibility of CPU mining bots, distributed denial of service bots, torrent bots, and more.

\subsection{Differences between WebGL and CPU Implementations} \label{implementation:challenges}
As stated in the introduction, WebGL is designed to provide graphics operations, so it is naturally more difficult to implement a GPU hash cracker than a CPU hash cracker. WebGL 1.0 has several limitations that create difficulties in implementing hash crackers. These include inability to return values other than pixel color, lack of dynamic access to arrays, no debugging abilities, lack of bitwise operators, and no $32$-bit support.  
All of these are needed to implement any cryptographic function. 
WebGL 2.0 makes the browser more vulnerable, as it introduces an improvement in its capabilities and relaxes some of the implementation limitations. It adds support for $32$-bit integers and provides implementation for the majority of bitwise operators (not all); however, the rest of the limitations are still present.

GPUs are massively parallel, with hundreds (if not thousands) of stream processors that can simultaneously calculate hashes. Although a CPU core is much faster than a GPU core, the CPU usually limited to only four to eight cores.

Hash cracking is highly suited to parallel computing due to the need to execute the same cryptographic functions on independent data sets. This gives GPUs a tremendous edge in hash cracking over CPUs. 
The open question we address is whether overcoming WebGL limitations would inflict a heavy cost on performance and eliminate the GPU hardware advantage over the CPU.

\section{Related Work}
\label{related:sec}
The feasibility of performing stealthy calculations using HTML5 Web Workers is presented by Rushanan ~\cite{rushanan2016malloryworker}. Web Workers are JavaScripts that run in the background, allowing web applications to spawn background workers in parallel to the main thread. Namely, it is possible to perform calculations without blocking the main thread, leaving the web pages’ UI loading time unaffected.
After the initial loading time, which takes about $3$ seconds, Rushanan ~\cite{rushanan2016malloryworker} managed to calculate 500K MD5 hashes per second.
Although the web page loading time seems unaffected, there is a rise in the CPU load. The rise in the CPU load might be noticeable if the user performs CPU intensive processes during the attack.

The first attempt to use WebGL 1.0 for distributed hash cracking was presented by MWR Labs \cite{MWRLabs:2012}. Even after overcoming the hurdles involved, which included packing input into textures, computing using a shader, and retrieving output from images, it became clear to them that this method of attack was not feasible.
The authors \cite{MWRLabs:2012} indicated that this was related to limitations in the WebGL 1.0 shading language, especially the lack of support for $32$-bit integers and bitwise operations.

One year later, Marc Blanchou presented his efforts to overcome both of the above challenges \cite{GPU:blackhat13}. He tried to use a vector with two floats and implemented the bitwise operations. However, this attempt was not an efficient implementation. Similar to the previous work cited \cite{MWRLabs:2012}, he concluded that the use of WebGL 1.0 for hash cracking was not cost effective. Blanchou also presented a comparison to a simple JavaScript CPU hash cracker, and showed that it would be faster. Furthermore, he added that the upcoming support of OpenGL ES 3.0 (upon which WebGL 2.0 is based) is an upgrade and would not have the limitations of WebGL 1.0.

Recently, a new phenomenon known as cryptojacking was discovered. This involves the in-browser mining of crypto-currencies, sometimes even after the browser window is closed, \cite{CPUKiller:CheckPoint,Miners:PaloAlto,Extension:Mining,cryptojacking:csoonline,Coinhive:persistent}, and more. Specifically, this entails the mining of Monero\cite{Monero:Home} using Coinhive\cite{CoinHive:krebson,CoinHive} or similar JavaScript CPU mining tools. 
Several papers in the past year performed a comprehensive analysis and in-depth study of cryptojacking ~\cite{Cryptomining:RWTH,Cryptojacking:Concordia,Cryptomining:BackShot,Cryptomining:Minesweeper}. These papers examine and analyze the phenomenon and its prevalence, each in its own unique way. 
To overcome the naive detection methods, modern mining tools commonly use evasion techniques such as limiting CPU usage, code obfuscation, and hiding the malicious code in popular third-party libraries. 
Both Hong et al. ~\cite{Cryptomining:BackShot} and Konoth et al. ~\cite{Cryptomining:Minesweeper} introduce countermeasures that enable the user to detect the different in-browser CPU mining tools more efficiently than the existing naive methods. 
Some websites may use such mining tools as an alternative to ad-based financing or offer premium content in exchange for mining. 
Other websites unknowingly fall victim to attacks that cause them to unwittingly serve mining code that uses the computer resources of its visitors.

Both Hong et al. ~\cite{Cryptomining:BackShot} and Konoth et al. ~\cite{Cryptomining:Minesweeper} show that detecting miners by means of blacklists, string patterns, or CPU throttling alone is an ineffective strategy, because of both false positives and false negatives.  They thoroughly explored the mining attack structure, miner communication, and distribution methods of current in-browser CPU mining tools and provided important insights that allowed them to implement more efficient defense mechanisms. 
They prove that it is more effective to use their suggested behavioral-based detection methods by using either static analysis~\cite{Cryptomining:Minesweeper} or runtime profilers~\cite{Cryptomining:BackShot}.

Both Eskandari et al. ~\cite{Cryptojacking:Concordia} and Rüth et al. ~\cite{Cryptomining:RWTH} tried to investigate how often cryptojacking occurs in websites using two straightforward approaches. Eskandari et al. ~\cite{Cryptojacking:Concordia} queried the top million sites indexed by Zmap.io and PublicWWW.com to determine which websites contained coinhive.min.js script in their body. Over $30,000$ websites were found. Rüth et al. ~\cite{Cryptomining:RWTH} inspected the .com/.net/.org and Alexa Top 1M domains for mining code, and found mining code in a relatively low percentage 0.08\% of the probed sites; however, this still accounts for more than 100,000 websites. 

Konoth et al. ~\cite{Cryptomining:Minesweeper} crawled Alexa’s Top 1 Million websites for a week. Using static analysis, they managed to detect 1,735 websites containing cryptojacking code out of 991,513 websites in total, meaning 0.18\%.

On the other hand, Hong et al. ~\cite{Cryptomining:BackShot} used their CMTracker detector, which has two runtime behaviour-based profilers, to collect 2,770 cryptojacking samples from 853,936 popular web pages, including 868 among the top 100K in Alexa’s list, meaning 0.32\%.
In addition, according to their findings, 53.9\% of these identified samples would have not been identified with current widely used detectors that are based on blacklists.

The increase use of crypto-currency as an alternative means of payment and the rise in performance and compute resources provided by in-browser coding, specifically with the use of WebAssembly, have made cryptojacking very appealing to criminals as a continuous source of income.

We predict that cryptojacking has the potential to be very profitable in high traffic websites, but the potential harm to users introduces ethical problems that must be considered. Some of the problems include higher energy bills, accelerated device degradation, slower system performance, and poor web experience.  
This forecast led researchers ~\cite{Cryptomining:BackShot,Cryptomining:Minesweeper} to address the magnitude of the potential harm and investigate potential defense mechanisms against this type of CPU-based in-browser cryptojacking.
Some researchers state that the trust model of web, which considers web publishers as trusted and allows them to execute code on the client-side without any restrictions is flawed and needs reconsideration~\cite{Cryptomining:Puppets}.
Furthermore, it is essential to explore other means of in-browser cryptojacking that may attract attackers, alongside effective defense mechanisms where necessary. This increases the importance of our WebGL research.

\section{Password Cracking Attack}
\label{passwordcracking:sec}
Although researchers previously implemented hash cracking using WebGL 1.0, they haven't optimized it thoroughly, nor compared hash cracking over WebGL 1.0 to WebGL 2.0.
This section reviews the challenges WebGL 1.0 and WebGL 2.0 present in implementing hash functions. It also briefly describes the implementations of password cracking using WebGL 1.0 and WebGL 2.0.

Our work presents new optimizations that improve previous results \cite{MWRLabs:2012,GPU:blackhat13}. 
We show that password cracking using WebGL 1.0 can be as efficient as the exploit of CPU in computational aspects, while WebGL 2.0 outperforms the CPU. 
For the sake of comparison to previous works, the first algorithm we evaluate is MD5. 
We also introduce the improvements WebGL 2.0 brings and compare them to WebGL 1.0 and CPU results.

\subsection{Implementing MD5 in WebGL}
The MD5 \cite{MD5:RFC} hashing computation conflics with the WebGL limitations presented in Section \ref{implementation:challenges}.
The MD5 hashing algorithm processes a variable-length message into a fixed-length output of $128$-bits, while WebGL’s maximum output length is $32$-bits.
The algorithm consists primarily of bitwise operations operating on $32$-bit variables, while WebGL 1.0 doesn't support $32$-bit variables. 
These bitwise operations include rotate, XOR, AND, OR, and NOT. WebGL 1.0 doesn't implement bitwise operators at all, and WebGL 2.0 is still missing some of them.
This means we had to implement the missing bitwise operators ourselves in WebGL, and define a new unit for WebGL 1.0 to support $32$-bit calculations. We also had to efficiently divide the hashing process to adjust it for the reduced output size.

For the password cracking attack, the MD5 hashing algorithm needs to be implemented in WebGL while overcoming all the presented challenges. Unlike WebGL 1.0, WebGL 2.0 has fewer limitations and it would be easier to implement the hashing algorithm. Therefore, we expect WebGL 2.0 to provide a significant performance boost as a result of these improvements.
    
\subsection{Evaluation} \label{passwordcrack:evaluation}
Implementing a simple straightforward brute-force MD5 hasher on WebGL is not enough. There are several algorithmic improvements and optimizations that must be evaluated and considered. These optimizations should use multi-threading effectively and maximize the number of calculations done per each draw call.

Furthermore, the evaluation done by Rushanan \cite{rushanan2016malloryworker} was performed on an old i5 processor and therefore the results are obsolete. It was essential to get more up-to-date results on how an MD5 hasher performs on a browser using a modern CPU, and to compare these results as well.

Based on a GitHub project of an MD5 brute-force password cracker \cite{MD5:CRACKER}, we implemented a JavaScript password cracker we could evaluate on a modern CPU. 

\paragraph{Evaluation Experiment}
To evaluate the proposed algorithms we devised an experiment on our personal computer to compare the results of WebGL 1.0, WebGL 2.0, and CPU-based techniques.
We tested several numbers of threads and HTML workers \cite{webworkers} for both WebGL 2.0 and CPU password crackers. For the WebGL 2.0 miner, our parameters are calculations per thread and the number of threads that combined give us calculations per draw. For the CPU miner, we were able to control the number of HTML workers. We ran each combination in turn for a duration of two minutes. During each one of the combinations, we measured the load on the GPU or CPU accordingly.
This allowed us to find the best trade-off values to use, while keeping the highest possible hash rate that would not reduce performance for the user.
Before the evaluation process, we tested several parameters for WebGL 1.0 password cracker and found the best trade-off values as well.

We performed one experiment on password cracking as an intro to in-browser attacks in general, and more specifically WebGL attacks. Our goal was to test how well the WebGL password cracker performed compared to a CPU-based cracker. For this test, we didn't evaluate large scale experiments that challenge other aspects of distributed attacks, such as: user experience, stealthiness, and possible defenses.

We did, however, evaluate large scale extended experiments on cryptocurrency mining attacks, as detailed in Section \ref{crypto:sec}. Password cracking attacks don't pose as viable a threat as cryptojacking, and are much less common than cryptocurrency mining attacks that use in-browser techniques.
\paragraph{Evaluation Setup} \label{Evaluation Setup}
Intel i5 6600K @3.5Ghz, 2X4GB @2400MHz, 
Radeon HD6870, Chrome Canary 64.0.3282

\subsubsection{Experiment 1 : Password Cracking}

\paragraph{Goal:} Compare the effectiveness of password cracking using WebGL 1.0, WebGL 2.0, and CPU techniques.
Also examine when the user starts to notice a password cracking process running in the background while using any of the above techniques.

\paragraph{Process:} We ran each of the proposed techniques in turn for a duration of two minutes for each parameter, while the user continued using the web-browser and other programs to simulate practical conditions. We then compared the hash rate achieved by each of the techniques and checked with the user when they noticed any impact on their usual experience.

\paragraph{Results:}

The best trade-off parameters for the evaluation were the ones that gave us the highest hash rate, while keeping the load around 50\%. Although better hash rates than the chosen results in Table 1 can be achieved in WebGL and CPU, these are the best trade-off values that did not affect the user experience.

\begin{figure}[!ht]
  \includegraphics[width=1\textwidth]
    {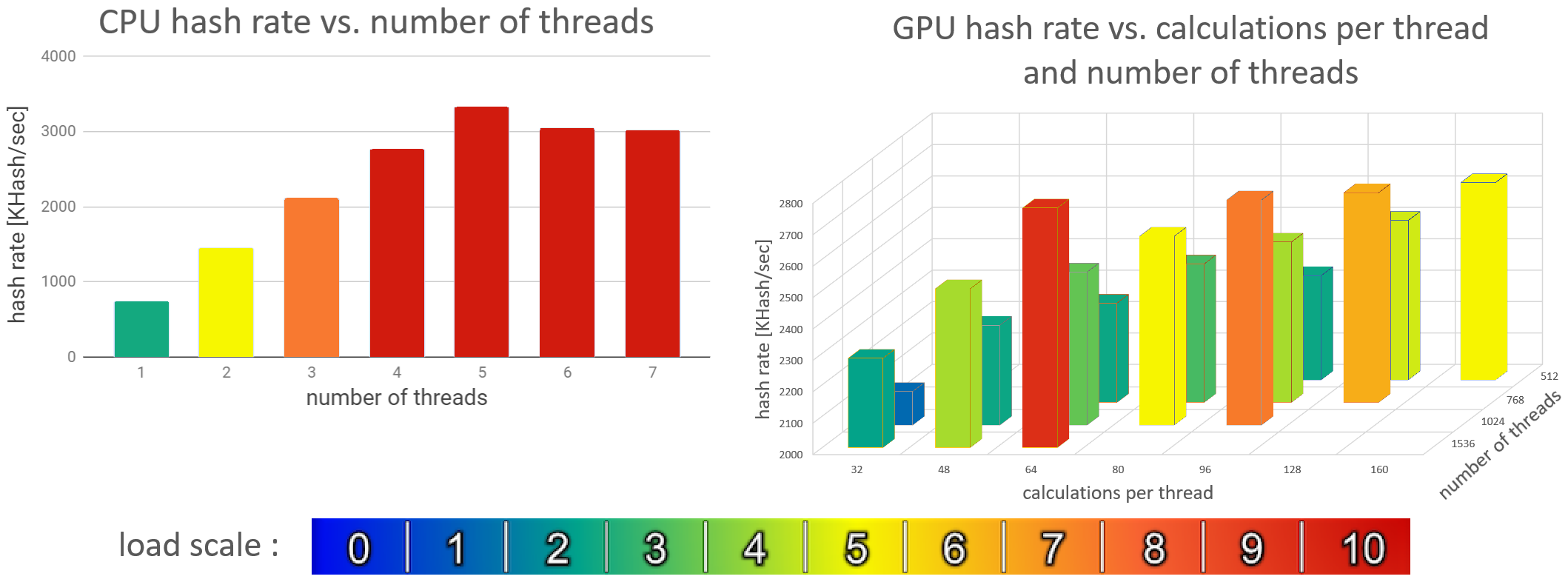}
  \caption{Password Cracking - Experiment 1 Results}
  Each column in both graphs presents the hash rate achieved and it's corresponding parameter.
  Each column is colored according to the measured percentage load.
\end{figure}

\begin{table}[!ht]
\begin{center}
\caption{Comparison between CPU, WebGL 1.0 and WebGL 2.0 hash rate results using best trade-off parameters}\label{tab10}
 \begin{tabular}{||p{6em} | p{3.5em} | p{5.3em} | p{5.5em} | p{4.5em}||} 
 \hline
 & threads & calculations\newline per thread & Hashes in\newline 10 minutes & Rate [hash/sec]\\ [0.5ex] 
 \hline\hline  
 WebGL 1.0 & 256 & 16 & 65212416 & 108687 \\ 
 \hline
 WebGL 2.0 & 512 & 160 & 1577615360 & 2629359\\ 
  \hline
 CPU & 2 & NA & 870187800 & 1459854\\ [1ex] 
 \hline
\end{tabular}
\end{center}
\end{table}

\noindent We can see that even our best result in WebGL 1.0 doesn't perform as well as the hash rate we managed to achieve on a modern CPU. 
However, there is a significant improvement when using WebGL 2.0 compared to WebGL 1.0; it even outperforms the hash rate achieved in our CPU hasher.

As part of the evaluation, we also combined both WebGL and CPU hashers to run simultaneously. This allowed us to gain the full benefit of both GPU and CPU due to the fact that each one of them uses different hardware resources.

This experiment shows the effectiveness of WebGL-based password cracking attacks. 

\section{Cryptocurrency Mining Attack}
\label{crypto:sec}
The rising popularity of both purchasing and mining cryptocurrency, caused a significant growth of the mining community and the introduction of new cryptocurrencies. 
This section introduces the two cryptocurrencies we experimented with: Monero \cite{Monero:Home} and Litecoin \cite{LITECOIN:HOME}. For Monero, we evaluated the effectiveness of WebGL in distributed cryptocurrency mining and compared it to the JavaScript CPU miner, Coinhive\cite{CoinHive}. While for cryptomining Litecoin, we only did an initial WebGL evaluation and compared it to a native GPU miner as additional studies. This seemed logical since Monero is more popular than Litecoin. 

As noted in Section \ref{Structure}, we wanted to compare the miner part of the cryptocurrency mining attack. We do not address the performance of other aspects in distributed cryptocurrency mining, such as mining pools, server side, databases, and more. For these elements we used commercial tools to which we added our WebGL miner; we only measured the mining performance.

\subsection{Monero} \label{Monero:intro}
A new type of cryptocurrency mining was introduced: CPU mining using JavaScript-based miners \cite{CPUKiller:CheckPoint}.
These JavaScript mining tools can be injected into popular websites as a source of income, at the expense of users’ resources.
The most common cryptocurrency mined using these web browser mining tools was Monero \cite{Monero:Home}. 
Monero experienced rapid growth in market capitalization and transaction volume during the year 2016. This was partly due to its adoption by the major darknet market AlphaBay, which was later closed down in mid-2017 \cite{Monero:Darknet,Monero:Hist}. 

Monero’s proof of work, the process that must be done to ensure a block is valid before it is added to the blockchain, is based on the underlying CryptoNote protocol \cite{CryptoNote}, called CryptoNight \cite{CryptoNight:RFC}. Unlike Litecoin and other cryptocurrencies that are derivatives of Bitcoin, the CryptoNight proof-of-work hash algorithm is a memory-intense function. Monero's proof-of-work algorithm is designed to be inefficiently computable on GPU, FPGA, and ASIC architectures, which makes it ideal for mining on CPUs.
Therefore, the two main features of the algorithm that challenge our WebGL implementation are:
\begin{description}
\item[$\bullet$] CryptoNight uses large fast memory to work on 2MB (L2 cache size), which requires a lot of silicon. This is far more than what is needed by the SHA-256 circuitry used for Bitcoin, Litecoin, and other similar cryptocurrency mining algorithms.
\item[$\bullet$] CryptoNight is based on AES (Advanced Encryption Standards, a cryptographic cipher applied by the majority of organizations) \cite{AES:NIST} and was designed to take advantage of the AES-NI instruction set \cite{AES-NI:INTEL}, which uses existing hardware circuitry on modern x86\_64 CPUs to speed up AES operations.  
\end{description}

The following section describes the challenges presented in implementing the CryptoNight hashing algorithm in WebGL. 
\subsection{WebGL Monero Implementation Challenges}
Implementing cryptocurrency mining on WebGL 1.0 has proven to be inefficient. The performance cost is too high since the use of bitwise operators and $32$-bit variables (which we implemented ourselves) is significantly greater compared to the MD5 hashing algorithm. Therefore, we evaluated cryptocurrency mining experiments only with WebGL 2.0.

After the initialization of an AES key from the input using the Keccak hashing algorithm \cite{Keccak:Offical}, the CryptoNight algorithm consists of three main steps: initializing a 2 MB scratch pad, executing a memory-hard loop, and finalizing the hash output. Each step presents different implementation challenges.

\paragraph{WebGL’s memory limit.} WebGL is unable to allocate a consecutive 2MB memory array to be used as a scratch pad for the first step. We could split the array into smaller chunks, but it would significantly affect the algorithm's performance.
\paragraph{Potentially long runtime.} After initializing the scratch pad, a memory-hard loop of 524288 iterations on non-consecutive array elements is performed; this can take a significant amount of time. 
WebGL calls are done in the main UI thread, so we want to minimize the time it takes for a WebGL call to return. Long periods would cause the users to feel the page was unresponsive or even lead to a context-lost event for WebGL.
\paragraph{Large shader code.} For the CryptoNight output, a hash function is randomly chosen out of four possibilities and applied on the state resulting from the previous steps. The resulting hash is the output of CryptoNight's algorithm. This step presents us with the new challenge of implementing all four possible hash functions: Blake \cite{BLAKE2:RFC}, Groestl \cite{GROESTL:SITE}, JH \cite{JH:SITE}, and Skein \cite{SKEIN:SITE}. This would result in a major increase of the shader code size.

Finally, as we stated before, GPUs are all about parallelism. We needed to make the CryptoNight algorithm parallel, despite the fact that there is no natural way to share data between shader threads.

The question remains whether overcoming the challenges in implementing the CryptoNight algorithm on WebGL would result in a severe cost in performance and make WebGL unusable for Monero mining.

\subsection{Experiment 2: Cryptomining Performance and User Experience} \label{exp2}
We overcame the major challenges presented in each step of the CryptoNight algorithm by using data textures, dividing the data efficiently, and moving some of the large code but non-intensive calculations to the CPU side.
After implementing Monero's algorithm in WebGL 2.0, we can proceed to evaluating the major aspects of the Monero mining attack.

In our experiment, we compared our WebGL 2.0 miner to the Coinhive browser-based CPU miner introduced in Section \ref{passwordcracking:sec}.
The results compared the performance, user experience, and hardware load locally on our computer.

\paragraph{Evaluation Setup} \label{Evaluation Setup Crypto}
Intel i5 6600K @3.5Ghz, 2X4GB @2400MHz, 
GeForce GTX1080, Chrome Canary 64.0.3282 \newline 

Similar to evaluating the password cracking attack, we tested different parameters to determine how the GPU and CPU load are affected. This allowed us to find the best trade-off values for a cryptocurrency mining attack.
To evaluate the proposed algorithms, we devised an experiment to compare the results of WebGL 2.0 and CPU-based techniques.
We tested several parameters for both WebGL 2.0 and Coinhive Monero miners. 
For the WebGL 2.0 miner our parameters were: number of threads and time between draws. Shorter times between draw calls means a higher hash rate, but it also leads to a higher GPU load.
WebGL 2.0 did not allow us to choose more than 32 threads.

For the CPU miner, we controlled the number of HTML workers \cite{webworkers} in Coinhive.
We ran each parameter combination of the GPU and CPU miners in turn for the duration of two minutes. During each of the combinations, we measured the hash rate and the load on the GPU or CPU, accordingly.
\paragraph{Goal:} Find the best trade-off evaluation parameters and compare the effectiveness of Monero cryptocurrency mining using WebGL 2.0 and CPU techniques.

\paragraph{Process:} We ran each of the proposed techniques and parameters in turn for a duration of two minutes while the user continued using the web-browser and other programs to simulate practical conditions. We then compared the hash rate achieved by each of the techniques and checked with the user when they felt any effect on their usual experience. 

\paragraph{Results:} The best trade-off parameters for the evaluation were the ones that gave us the highest hash rate, while still not affecting the user experience.

\begin{figure}[!ht]
  \includegraphics[width=1\textwidth]
    {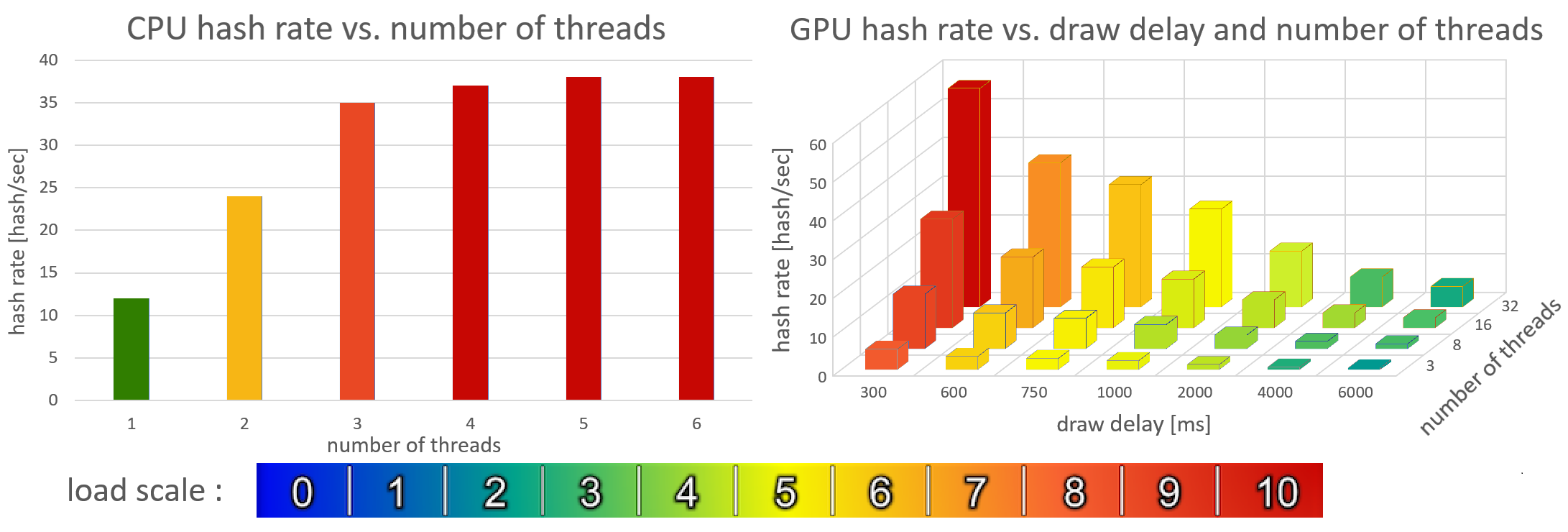}
  \caption{Cryptomining - Experiment 2 Results}
  Each column in both graphs presents the hash rate achieved and it's corresponding parameter.
  Each column is colored according to the measured percentage load.
\end{figure}

We can see that the best trade-off results would be achieved if we limit Coinhive to using 2 threads, which leads to about 60\% CPU usage. For the WebGL 2.0 miner, the best performance was achieved using 32 threads with  a 1 second draw delay, resulting in 50\% GPU usage.
Using these best trade-off parameters, we reached a hash rate of 25 hashes/sec in WebGL and 24 hashes/sec in Coinhive. We observed that the WebGL miner is just as fast as the Coinhive miner. We did not detect any impact on the user experience when we evaluated performance on a single user's machine.

\subsubsection{Litecoin}
As additional studies we decided to test whether WebGL mining is efficient with another type of cryptocurrency, called Litecoin \cite{LITECOIN:HOME}. Litecoin gained a lot of popularity and shows great promise. It's mining difficulty is lower than that of Bitcoin \cite{BITCOIN:PAPER}, and the market is not yet saturated with dedicated ASIC miners.
Litecoin is a fork of the Bitcoin core client. It differs primarily in requiring less time to mine a block (2 minutes instead of 10 minutes), a higher maximum number of coins, and different hashing algorithm: Scrypt \cite{SCRYPT:RFC} instead of SHA-256 ~\cite{SHA256:RFC}.

\paragraph{Implementation and Challenges}
Cryptocurrencies that use Scrypt are often mined on GPUs, which tend to have significantly more processing power (for some algorithms) compared to the CPU. 
Although at first the Scrypt algorithm seems to be complicated, implementing Scrypt in WebGL 2.0 is relatively straightforward because it is suitable for GPU computation.
The main difficulties in implementing Scrypt in WebGL 2.0 involve: 
\begin{description}
\item[$\bullet$] Splitting the algorithm to allow the dynamic hash output to be divided between the $32$-bit pixel output from each shader.
\item[$\bullet$] Implementing PBKDF2(HMAC-SHA256) \cite{PBKDF2:RFC}, which includes SHA-256 implementation with different key lengths, while keeping the shader code small enough so we don't reach the maximum number of instructions per shader program.
\end{description}

\paragraph{Evaluation} The initial hash rate we managed to achieve with WebGL 2.0 was around 210 kHash/sec, which is about half the hash rate a native GPU miner (OpenCL\cite{Khronos:OpenCL}) can reach with the same hardware \cite{Mining:BM,Github:Hashcat}.

\section{Evaluating the Distributed Attack} \label{crypto_dist:sec}
The previous two sections analyzed the exploit of WebGL 2.0 from the narrow perspective of single user. An attacker who aims to abuse WebGL for password cracking or cryptocurrency mining, must launch the attack on many web users. 
This section analyzes the practical aspects of distributing these attacks.

Of the two attacks, we chose to evaluate the distributed attack on cryptocurrency mining for two reasons:
\begin{enumerate}
    \item Distributed cryptocurrency mining attacks are more common today.
    \item We wanted to evaluate distributed attacks from a practical perspective. In cryptocurrency mining attacks, we have other real-world implementations of the attack that can be used for comparison.
\end{enumerate}

Due to the similarity between the attacks, we believe that the findings can also be used to reach the same conclusions for password cracking.

In Section \ref{Structure} we introduce cryptocurrency mining distributed attacks and our implementation. Section \ref{Distribution} discusses different aspects of distribution on a large scale and how we handled them.

In Section \ref{exp3:sec} the experiment evaluates the user experience of the attack using our WebGL miner and Coinhive with specific parameters.
Section \ref{exp4:sec} describes our experiment to check the stealthiness of the attack. We wanted to examine whether users could locate the attack if they are aware it is running on their computer.
Each of the experiments focuses on a different aspect of distributed cryptocurrency mining attacks, while providing additional data about the mining hash rate.

\subsection{Implementing the Distributed Attack} \label{Structure}
Similar to what we introduced in \ref{preliminaries:sec}, the goal of distributed attacks for cryptocurrency mining is to find input data for which, after applying the appropriate hashing function, the result matches the attacker's desired criterion: it needs to be less than the chosen cryptocurrency’s current target value. This would give the mining reward to the attacker based on work done using the victims’ resources. The attack is performed by dividing the data search among as many clients as possible.

The process of distributed hash cracking attacks on the victim's browser works as follows: 
\begin{enumerate}
\item Malicious code reaches a victim as per the distribution methods described in Section \ref{Distribution}
\item Attacker's code in the victim's browser sends a message to the attack server asking for a target and data range 
\item The code calculates the appropriate hash function on the data range using the victim's resources
\item The code compares the crypto hash function's output to the desired target value
\item The results are posted from the victim's browser to the attack server
\item The server sends a new target and data range to run on the victim's browser 
\item The process returns to Step 3 until there are no calculations to be done for the attacker
\end{enumerate}

In addition to the attack process on the victims' side as described above, the cryptocurrency mining attack involves several additional elements that an attacker needs to address.
Cryptocurrency mining starts by creating a new mining pool or joining an existing one. The mining pool is a group of cooperating miners who agree to share block rewards in proportion to the mining hash power they contribute.
After the server has a working mining pool, it is ready to send mining tasks to its member clients.

As stated in Section \ref{passwordcracking:sec}, cryptocurrency mining is a race to find the corresponding hash, so time is of essence.
Performance and speed play a vital part in the success of cryptocurrency mining. Even if we are behind the competitors by just a fraction from posting the correct hash, this will mean we miss out on the mining reward.
In a distributed cryptocurrency mining attack, the server can divide the data range between its cooperating miners and increase the chances of finding a match to the transaction hash. The potential reward increases with the computational speed.

In the following experiments we evaluated Step 3 of the distributed hash cracking process (described above) using WebGL calculations. We show the effectiveness of WebGL in mining Monero cryptocurrency as compared to an equivalent CPU hasher. This should prove that WebGL can handle mining cryptocurrency that is of significant relevance today. 

\subsection{Large Scale Distribution Aspects} \label{Distribution}
In this section we describe the different issues an attacker needs to consider before launching a large-scale distributed attack, and how we resolved them for our evaluations.

The browser's GPU can be abused by an attacker to conduct efficient distributed password cracking attacks, crypto-currency mining, and more.
To start spreading the attack, the attacker just needs to somehow inject their JavaScript code into a website that will reach as many users as possible. An attacker can achieve this through several methods, depending on her technical knowledge and resources:
\begin{description}
\item[$\bullet$] Come to an arrangement with websites to insert attacker code, for example, by sharing earnings
\item[$\bullet$] Pay an ad company to pop ads containing attacker's code \cite{Dorsey:Medium}, this can be done with or without the ad company’s knowledge
\item[$\bullet$] Inject attacker code maliciously into websites
\item[$\bullet$] Come to an arrangement with extension companies, such as AdBlock\cite{AdBlock}
\item[$\bullet$] Develop a popular website to lure victims
\end{description} 

An attacker who wishes to launch such an attack on many web users for an extended period of time, must consider additional aspects of the attack:
\begin{enumerate}
\item {\em Stealthiness}. The attack must be conducted without arousing the user’s suspicion. If the user feels any impact on his browsing experience, he may close the website or even contact the website's owner. This can lead to detection and prevent the attack from running for a prolonged period of time.
\item {\em Management}. Assuming thousands of browsers run the attack for different periods of time, it is necessary to manage them all to maximize the profit. Duplicate runs need to be avoided because they waste valuable computing power. Moreover, there is a need to keep track of users going offline to prevent computations from being skipped. To accomplish this, the attacker needs an efficient and synchronous control server.
\end{enumerate}

\subsection{Implementing Distributed Attack Experiments}
In our experiments, we didn't need any of the distribution methods mentioned above because we had volunteers who knowingly entered websites that contained our attacking code. 

In Section \ref{crypto:sec}, for CPU mining experiments we used Coinhive, which handled the management of all the users. It also enabled some degree of stealthiness by setting the CPU usage limit to avoid detection and by the use of WASM, which makes it difficult to find the mining code.

For experimental purposes only, we implemented our own naive WebGL mining server to handle our user scale. Then we installed a NodeJS server \cite{NodeJS:Home} that served mining jobs to each of the clients. We used a MongoDB \cite{MongoDB:Home} to keep track of ongoing work and store the target hashes. To keep things simple and isolate our WebGL miner, we didn't connect it to any active Monero mining pools.

For the client side, we had an iframe with obfuscated code; this received the WebGL mining code from our server and contributed greatly to our miner's stealthiness. We also limited the WebGL performance, using the best trade-off parameters observed in Experiment 2 in Section \ref{exp2} to prevent high GPU usage, with a minimal effect on the user experience. 

As stated in Section \ref{Structure}, we only planned to evaluate the effectiveness and performance of Step 3 of the in-browser attack process. Consequently, we used commercial tools for the other steps to narrow our measurements to the achieved hash rate and user experience. 

\subsection{Experiment 3: Distributed Cryptocurrency Mining Performance and User Experience} \label{exp3:sec}
The following distributed experiment extended the previous one described in Section \ref{exp2} to test several chosen best trade-off parameters on a larger scale. Our GPU was relatively high end and to see how well we could perform on weaker GPUs, we also used some of the parameters that showed only 30\% load on our GPU.
This enabled us to show the relevance of our local results to a wider range of GPUs and CPUs. 
Similar to previous experiments, we ran each combination in turn for two minutes. During each of the combinations, the user was asked to state whether he felt any effect on his computer’s performance. 
\paragraph{Goal:} Check whether the user notices the effect of a cryptocurrency mining process running in the background. Further compare the effectiveness of distributed cryptocurrency mining using WebGL 2.0 and CPU techniques.

\paragraph{Methodology and ethics:} The experiment was carried out with 100 volunteers. All the volunteers were paid to cover their expenses (primarily electricity), signed a consent form, and used their own computers to simulate a more realistic scenario. Mining can result in high electricity use and even lead to physical damage to less suitable devices (overheating cellphones for instance), so we only ran the experiment on personal computers. To avoid unintentional bias by participants and/or staff, the experiment was 'double blinded'. The users were assigned randomly to one of two sets, (\textit{CPU mining} or \textit{GPU mining}), without either the user of the staff being aware of the assignment.
We didn't collect any statistics or personal information from the volunteers so there were no privacy issues.

\paragraph{Process:} Users were asked to visit the website that contained our cryptocurrency mining logic and leave it open to run in the background. The users were then instructed to continue using the web-browser and other programs to simulate practical conditions for the duration of 10 minutes, and write down the times when they felt any influence on their browser’s performance. There were three possible answers: (1) Significant effect. (2) Minor effect. (3) No effect. We then randomly assigned to each visitor one of the two mining options: \textit{CPU mining} or \textit{GPU mining}. For \textit{CPU mining} users, we ran Coinhive Wasm miner in the background. For \textit{GPU mining} users, we ran our WebGL miner in the background. 

For each visitor, we saved the hash rate we managed to achieve with each parameter combination, and used it to compare between GPU and CPU mining hash rates.

\paragraph{Results:}
We observed that most of the users assigned to a CPU miner did notice some effect on their browsing experience.
\begin{figure}[!ht]
  \includegraphics[width=1\textwidth]
    {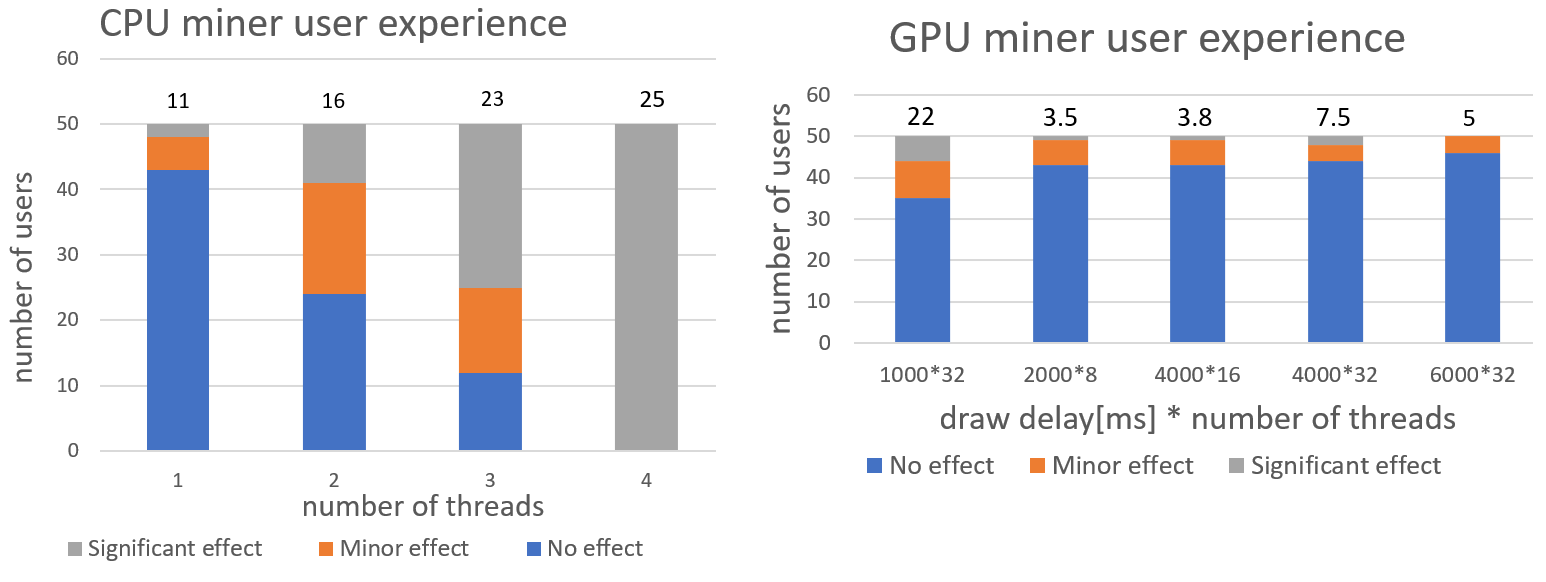}
  \caption{Distributed Cryptomining - Experiment 3 Results}
  The results show the degree to which users felt their computer’s performance was affected, if at all. Above each column is the average hash rate [hash/sec].
\end{figure}

A substantial number of users have CPU usage monitoring tools that allowed them to immediately notice the browser's high CPU use.
Moreover, some of the users were simultaneously running CPU intensive tasks in the background, which resulted in a significant deterioration of their computer's performance.

On the other hand, only a few users out of the GPU mining group reported noticing any effect. 
The common ground for the users that did feel some deterioration in the GPU group was that it occurred during the period with a high number of threads and a low draw delay. Moreover, their computers were relatively old and weak, in particular their GPU (more than five years old) or integrated GPUs. 

We can see that the best trade-off parameters from the previous experiment gave us a high average hash rate of 22 hashes/sec. 
However, almost a third of the users did notice some effect on their performance when using this parameter combination of 32 threads with a 1 second draw delay.
Perhaps this could have been avoided if we had added adaptation logic in our WebGL 2.0 miner to make it reduce its performance on weaker machines. Until we do, in our opinion, it is better to choose less demanding parameters for a wide range attack.

The best trade-off values for WebGL 2.0 that give us the highest hash rate while having a minimal effect on user experience (about 10\%) are: 32 threads and a 4 second draw delay.
This gave us the average hash rate of 7.5 hashes/sec for the WebGL 2.0 miner, while the Coinhive miner’s average hash rate reached 16 hashes/sec with 2 threads.
That said, if we had demanded the same degree of minimal effect on user experience from the CPU miner, we would need to limit Coinhive to using only 1 thread, giving us an average hash rate of 11 hashes/sec.
We concluded that, although we could reach a higher hash rate in WebGL 2.0, we should compromise on lower values until we optimize the miner with adaptation logic. 

This experiment demonstrated that the CPU hash rate is faster than the GPU hash rate in typical home machines.
The results also indicate that the user experience is affected much more during CPU attacks than GPU attacks, making the CPU attacks easier to notice and detect. Sixty percent of the CPU miner group reported degraded performance compared to only 15\% on average in the GPU miner group.
This corroborates our view that the GPU can be used for effective attacks. 

The fact that users notice an intensive task happening in the background doesn’t necessarily mean that they can find it. We therefore conducted Experiment 4 to evaluate how difficult it would be for them to discover the attack in each miner type (CPU or GPU).

\subsection{Experiment 4: Cryptocurrency Mining Stealthiness}\label{exp4:sec}

\paragraph{Goal:} Using the best trade-off evaluation parameters from previous experiments, observe whether the user can locate our cryptocurrency mining code running in the background.

\paragraph{Methodology and ethics:} Similar to the previous experiment to measure the user experience during cryptocurrency mining, the experiment was carried out with 80 volunteers. The volunteers were paid to cover their expenses (primarily electricity), signed a consent form, and used their own computers to simulate a realistic scenario. Mining could lead to physical damage on less suitable devices (overheating cellphones for instance), so we only ran the experiment on their personal computers. To avoid unintentional bias by participants and/or staff, the experiment was 'double blinded.' The users were randomly assigned to one of two sets, \textit{CPU mining} or \textit{GPU mining}, without either the user or staff aware of the assignment.
The users were told that we are going to try to steal money from them and they need to find out how.
We didn't collect any statistics or personal information from the visitors so there were no privacy issues.

\paragraph{Process:} Each user received a list of five different websites. One of them contained our mining logic from the previous experiment; the other websites were 'clean'. The users were instructed to open all of the websites and leave them open in the background.
We then randomly assigned each visitor to one of the two mining options: \textit{CPU mining} or \textit{GPU mining}. For \textit{CPU mining} users, we ran Coinhive miner in the background. For \textit{GPU mining} users, we ran our WebGL miner in the background. The users had to detect how we were stealing money from them i.e., find our CPU miner or GPU miner (depending what they were assigned).

There were four possible outcomes to this experiment: 1) Found \textit{CPU mining}, 2) Found \textit{GPU mining}, 3) Didn't find \textit{CPU mining}, 4) Didn't find \textit{GPU mining}.

\paragraph{Results:}
We observed that when it comes to stealthiness, GPU-based miners are much more effective and harder to detect than CPU-based miners.
In the CPU group, 22 out of 40 users were able to find the CPU mining code, representing 55\% of the users. 
Out of 40 users in the GPU group, only 1 user with a high level of technical knowledge in web research was able to find the GPU mining code, representing only 2.5\%.
Clearly, an attacker who launches a GPU-based attack on a large scale can be less concerned about detection. 
This strengthens the advantages of GPU miners as compared to CPU miners.

\section{Defenses}
\label{defenses:sec}
Today, WebGL is still not widely used by most of the common websites. Most do not harness the power of the GPU to perform any of the rendering on their pages. Apparently, WebGL is primarily being used in online web games or websites containing 3D imaging. Therefore, in our opinion, the most effective way to prevent WebGL attacks would be using an extension that can disable WebGL when it's not supposed to be used. 

Similar to web notifications \cite{Web:Notification}, which can send alerts to the user outside the context of a web page, it would be preferable to have WebGL disabled by default. The user would have the option to enable WebGL for each website individually, if it is required.

We considered looking into other directions for defense mechanisms to prevent the abuse of WebGL. For example, trying to detect when the canvas color is posted to a remote server, since it should only be used locally for further rendering. Another example might detect the increased use of bitwise operations inside shader code, which typically characterizes crypto calculations. However, we felt that this would be an overkill and we should consider a more simple solution.

Until a solution is addressed by browsers, our extension could operate as follows:

\begin{description}
\item[$\bullet$] Detect that the current website is using WebGL.
\item[$\bullet$] Alert the user that the current website is using WebGL.
\item[$\bullet$] Ask the user if the current website is supposed to use WebGL in any way, say for: online gaming, 3D imaging, augmented reality, complex geometric rendering, and more.
\item[$\bullet$] Allow the user to easily disable WebGL for the current website if none of the above conditions are met.
\end{description} 
We could also maintain a blacklist of websites for which WebGL should be disabled by default and reduce the need for user interaction.
This extension would enable us to minimize or even entirely eliminate WebGL attacks.

The following experiment allowed us to evaluate how well our extension works and to test our assumption that most of the websites don't use WebGL.

\subsection{Experiment 5: Extension Effectiveness}

\paragraph{Goal:} Collect statistics about the use of WebGL in websites and test the extension's efficiency.

\paragraph{Methodology and ethics:} The experiment was carried out with 50 volunteers who installed the extension on their personal computers for 3 weeks. They were instructed to keep their usual browsing habits and not try to test the extension intentionally. They were also asked to report if they noticed anything unusual: slow browsing, pages not loading, crashes, and so forth.
The extension reported back statistics on how many websites the user visited, how many of them contained WebGL code, and how many users chose to disable WebGL.
We didn't collect any personal information on the visitors so there were no privacy issues.

\paragraph{Process:} Each volunteer installed our extension and continued using their personal computer as usual. After three weeks, we instructed them to uninstall the extension and we reviewed the statistics collected by the extension.

\paragraph{Results:} 
None of the volunteers reported any issues regarding the extension.

During our evaluation, the extension encountered 1345 websites in total. Surprisingly, during the experiment period our extension came across very few websites containing WebGL. The extension encountered only 6 websites containing WebGL, and only one instance where the user chose to disable WebGL. We conjecture that some (if not all) of these WebGL website entries derive from the users wanting to challenge our extension, although they were instructed against this.

These results strengthen our assumption that WebGL websites are only a small fraction of existing sites and most websites don't use WebGL rendering. For the average user, disabling WebGL by default in their browsers, probably will not have any effect on their browsing experience and can only enhance their security.

\section{Conclusions}
\label{conclusions:sec}
Most of the popular web browsers today support WebGL 2.0 and it is enabled by default for all websites. WebGL is completely integrated into the web standards of these web browsers, allowing GPU-accelerated usage of image processing and 3D effects as part of the web page canvas. WebGL allows browsers to communicate directly with graphics hardware, enabling code to harness the GPU’s power. Currently, web browsers have not implemented any countermeasures or detection mechanisms for malicious WebGL code or the misuse of WebGL.

Our experiments show that WebGL 2.0 has introduced major improvements over WebGL 1.0 both in performance and in convenience. We also show that WebGL 2.0 can be used in practical attacks, such as exploitation for cryptocurrency mining, where in some cases it even outperforms CPU-based attacks. We demonstrated how difficult it is for a user to detect attacks that use GPU-based techniques compared to similar CPU-based techniques.
WebGL allows an attacker to benefit financially by abusing users' resources without their knowledge. Our work also suggests a practical defense mechanism.

We hope this paper will call attention to the problem and help tackle this vulnerability. It is our intention to raise awareness regarding the risk posed by WebGL, and the need for this risk to be addressed by web browsers.

\bibliographystyle{unsrt}

\end{document}